%% file: surfMG.tex
\documentclass[acmtog,balance = false]{acmart}

\acmJournal{TOG}
\setcitestyle{square}

\input{derekPre}

\setcopyright{rightsretained}
\acmJournal{TOG}
\acmYear{2021}\acmVolume{40}\acmNumber{4}\acmArticle{80}\acmMonth{8} \acmDOI{10.1145/3450626.3459768}



\begin{document}

\title{%
  Surface Multigrid via Intrinsic Prolongation%
}

\author{Hsueh-Ti Derek Liu}
\affiliation{
  \institution{University of Toronto}
  \country{Canada}}
\email{hsuehtil@cs.toronto.edu}
\author{Jiayi Eris Zhang}
\affiliation{
  \institution{University of Toronto}
  \country{Canada}}
\email{jiayieris.zhang@mail.utoronto.ca}
\author{Mirela Ben-Chen}
\affiliation{
  \institution{Technion - Israel Institute of Technology}
  \country{Israel}}
\email{mirela@cs.technion.ac.il}
\author{Alec Jacobson}
\affiliation{
  \institution{University of Toronto}
  \country{Canada}}
\email{jacobson@cs.toronto.edu} 

\begin{teaserfigure}
  \vspace{-5pt}
  \includegraphics[width=\linewidth]{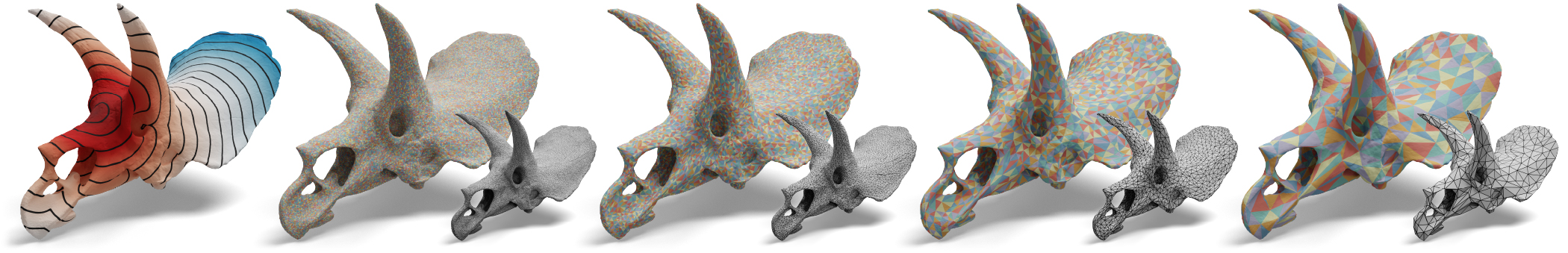}
  \vspace{-15pt}
  \caption{We introduce a novel geometric multigrid solver for curved surfaces.
  Our key ingredient is an \emph{intrinsic} prolongation operator computed via
  parameterizing the high resolution shape via its coarsened counterpart,
  visualized using colored triangles. By recursively applying this
  self-parameterization, we obtain a hierarchy (from left to right) for our multigrid method (e.g., to solve heat geodesics \cite{Crane2017HMD}, far left). \textcopyright model by Benoît Rogez under CC BY-NC.
  }
  \label{fig:teaser} 
\end{teaserfigure} 

\input{sections/abstract}

\maketitle
\input{sections/introduction.tex}
\input{sections/related.tex}

\input{sections/multigrid.tex}
\input{sections/prolongation.tex}

\input{sections/implementation.tex}
\input{sections/applications.tex}
\input{sections/discussion.tex}
\input{sections/conclusion.tex}

\bibliographystyle{ACM-Reference-Format}
\bibliography{sections/reference}
\input{sections/appendix.tex}

\end{document}

%% file: derekPre.tex
\usepackage{wrapfig}
\usepackage{nicefrac}
\usepackage{mathtools}
\usepackage{graphicx}
\usepackage{booktabs} 
\usepackage{combelow} 
\usepackage[ruled,vlined,linesnumbered]{algorithm2e}
\usepackage{tabularx} 
\usepackage{colortbl} 

\usepackage{manfnt} 

\usepackage{amsmath}
\usepackage{amsfonts}
\usepackage{amsbsy}

\usepackage[mathletters]{ucs}
\usepackage[utf8x]{inputenc}

\definecolor{derekBlue}{RGB}{144,210,236}
\definecolor{derekTableBlue}{RGB}{189,235,252}
\definecolor{iglGreen}{RGB}{153,203,67}
\definecolor{coralRed}{RGB}{250,114,104}
\definecolor{gray}{RGB}{180,180,180}
\definecolor{orange}{RGB}{255,165,0}
\definecolor{TechnionBlue}{RGB}{0,33,71}
\definecolor{lightgray}{gray}{0.65}

\SetCommentSty{commentFont}
\SetNlSty{textbf}{}{.}

\newcommand{\update}[1]{{#1}}

\newcommand{\refequ}[1] {Eq.~\eqref{equ:#1}}

\newcommand{\reffig}[1] {Fig.~\ref{fig:#1}}

\newcommand{\reftab}[1] {Table~\ref{tab:#1}}
\newcommand{\refsec}[1] {Sec.~\ref{sec:#1}}
\newcommand{\refapp}[1] {App.~\ref{app:#1}}
\newcommand{\refalg}[1] {Alg.~\ref{alg:#1}}

\DeclareMathOperator*{\argmin}{arg\,min}
\DeclareMathOperator*{\minimize}{minimize}

\newcommand{\R}{\mathbb{R}}

\newcommand{\M}{\mathcal{M}}
\newcommand{\N}{\mathcal{N}}
\def\arap{{\textsc{arap}}\xspace}
\def\lscm{{\textsc{lscm}}\xspace}

\newcommand{\vb}{\vu^{l}}

\newcommand{\ub}{{\vu}^{l}}
\newcommand{\ua}{{\vu}^{l+1}}

\renewcommand{\U}{{\mU}}
\newcommand{\Va}{V^{l+1}}
\newcommand{\Fa}{F^{l+1}}
\newcommand{\Vb}{V^l}
\newcommand{\Fb}{F^l}
\newcommand{\vecFont}[1]{\mathsf{#1}}

\def\vb{{\vecFont{b}}}
\def\vc{{\vecFont{c}}}

\def\vf{{\vecFont{f}}}

\def\vr{{\vecFont{r}}}

\def\vu{{\vecFont{u}}}
\def\vv{{\vecFont{v}}}

\def\vx{{\vecFont{x}}}

\newcommand{\matFont}[1]{\mathsf{#1}}
\def\mA{{\matFont{A}}}

\def\mF{{\matFont{F}}}

\def\mM{{\matFont{M}}}

\def\mP{{\matFont{P}}}
\def\mQ{{\matFont{Q}}}
\def\mR{{\matFont{R}}}

\def\mU{{\matFont{U}}}
\def\mV{{\matFont{V}}}

%% file: sections/abstract.tex
\begin{abstract}
  This paper introduces a novel geometric multigrid solver for unstructured curved
  surfaces. 
  Multigrid methods are highly efficient iterative methods for solving systems
  of linear equations. 
  Despite the success in solving problems defined on structured domains, generalizing multigrid to unstructured curved domains remains a challenging problem. 
  The critical missing ingredient is a prolongation operator to transfer functions across different multigrid levels. 
  We propose a novel method for computing the prolongation for triangulated surfaces based on intrinsic geometry, enabling an efficient geometric multigrid solver for curved surfaces. 
  \update{Our surface multigrid solver achieves better convergence than existing multigrid methods.}
  Compared to direct solvers, our solver is orders of magnitude faster. 
  We evaluate our method on many geometry processing applications and a wide variety of complex shapes with and without boundaries. 
  %
  By simply replacing the \update{direct solver}, we upgrade existing algorithms to interactive frame rates, and shift the computational bottleneck away from solving linear systems.
\end{abstract}

%% file: sections/introduction.tex
\section{Introduction}\label{sec:intro}
%
\begin{figure}
  \centering
  \includegraphics[width=3.33in]{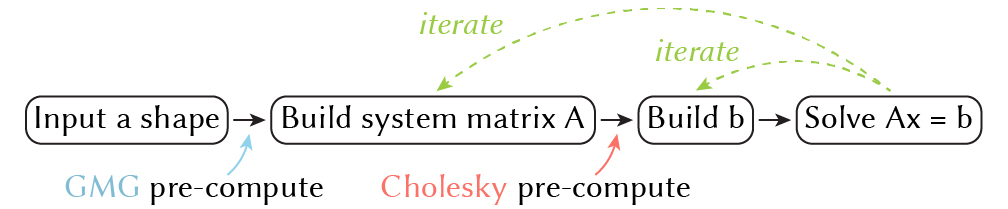}
  \caption{Many geometry processing algorithms that involve solving linear
  systems $\mA \vx = \vb$ often consist of four steps: (1) loading a geometry,
  (2) building \update{the} left-hand-side $\mA$, (3) building \update{the} right-hand-side $\vb$, and (4)
  solving the system $\mA\vx = \vb$. Direct solvers (e.g., Cholesky) perform
  pre-computation after building $\mA$, making it suitable for applications
  where only $\vb$ is changing. Geometric multigrid methods perform
  pre-computation solely based on the geometry. Thus, even the entire system
  $\mA, \vb$ \update{are} changing, geometric multigrid solvers can still leverage the
  same pre-computed hierarchy to solve the system efficiently.}
  \label{fig:fourSteps}
  \vspace{-5pt} 
\end{figure} 
Linear solvers are the heart of many geometry processing algorithms. For
positive (semi-)definite problems defined on surface meshes, direct solvers
(e.g., Cholesky factorization) are commonplace. Unfortunately, direct solvers do
not scale and often become the bottleneck for problems on high-resolution surface
meshes. Especially for applications where the linear system changes at
every iteration (e.g., simulation), direct solvers require an expensive
re-factorization.

For problems on structured domains (e.g., 2D/3D regular grids), an excellent
alternative is \emph{geometric multigrid} methods. Geometric multigrid solvers
perform pre-computation solely based on the geometry without knowing the linear
system of interest (see \reffig{fourSteps}). This enables multigrid methods to
solve the system efficiently in linear time even when the system changes at each
time step. 
Multigrid solvers already become non-trivial for unstructured grids (e.g., arbitrary
triangle meshes in 2D or tetrahedral meshes in 3D),
the added complexity of immersing triangle meshes in 3D
has left a ``black-box'' multigrid solver for 
curved surfaces elusive until now.



In this paper, we propose a \emph{Galerkin geometric multigrid} solver for
manifold surface meshes with or without boundaries. Our key ingredient is a
method for computing the \emph{prolongation} operator based on the intrinsic
geometry. Our multigrid solver achieves a better convergence rate compared to alternative multigrid methods. Replacing direct solvers with our black-box surface multigrid solver leads to orders of magnitude speed-up. We show our method is effective in a variety of applications ranging from data smoothing to shell simulation, with linear systems of different sparsity patterns and density. Our contributions turn existing algorithms into interactive applications (e.g.,
\reffig{smoothing_tuning}) and shift the bottleneck away from solving linear
systems (e.g., \reffig{mcf}). 


%

%% file: sections/related.tex
\section{Related Works}
Multigrid methods \cite{brandt1977multi} have earned a reputation as one of the
fastest numerical solvers for solving linear systems. On structured domains
(e.g., 2D/3D grid), multigrid is very well-studied both theoretically
\cite{trottenberg2000multigrid, hackbusch2013multi} and practically
\cite{brandt2011multigrid}. 
%
In graphics, multigrid has been an attractive solution for
interactive and large-scale applications on structured domains, most prominently
for image processing \cite{KazhdanH08, KrishnanS11} and simulating fluids on
large grids \cite{McAdamsST10,Aanjaneya2019,LaiCGBW20}.
Even for problems where the original representation is unstructured, an
auxiliary background grid can be introduced for multigrid to perform efficient
computation and transfer the solution back to the unstructured representation.
For example, one can run multigrid on a background hexahedral mesh to simulate
elastic deformations \cite{ZhuSTB10, DickGW11} and character skinning
\cite{McAdamsZSETTS11}. \citet{ChuangLBRK09} deploy multigrid on a background
voxel grid to solve Poisson problems defined on the surface mesh.
To reduce the complexity of using structured representations, adaptive multigrid
methods are developed for subsurface scattering \cite{HaberMBR05},
smoke simulation \cite{SetaluriABS14}, and other graphics applications
\cite{kazhdan2019adaptive}.

\paragraph{Unstructured Euclidean Domains}
Directly deploying multigrid to unstructured ``grids'' in Euclidean domains
(e.g., 2D triangle meshes and 3D tetrahedral meshes) has also been an important
problem for decades. The main difficulties lie in how to construct the multigrid
hierarchy and how to transfer signals back-and-forth across different grid
levels.
In graphics, unstructured multigrid for 2D triangle meshes is widely applied to
cloth simulation where the design pattern is prescribed by a 2D boundary curve.
In the methods proposed in \cite{OhNW08, JeonCKCK13,WangWFTW18}, they generate the hierarchy in a coarse-to-fine manner by
triangulating the 2D design pattern and then recursively subdividing it to get
finer resolutions. \citet{WangWFTW18} generate the multigrid hierarchy from
fine-to-coarse by clustering vertices on the fine mesh and re-triangulating the
2D domain. 
When it comes to 3D tetrahedral meshes, multigrid is commonly used to simulate
deformable objects. \citet{GeorgiiW06} build the hierarchy with different
tetrahedralizations of the same 3D domain. \citet{OtaduyGRG07} repetitively
compute the offset surface of the boundary mesh, decimate the offset surface, and
tetrahedral-mesh the interior to obtain the hierarchy. \citet{SachtVJ15} follow a similar technique but with more elaborate and tighter fitting offsets.
\citet{AdamsD99}
recursively remove the maximum independent set of vertices and \update{tetrahedralize} the interior. 
These unstructured multigrid methods for the Euclidean domains rely on the fact
that every level in the hierarchy is a triangulation or tetrahedralization of
the same space. Thus, they can easily define linear prolongation using
barycentric coordinates on triangles or tetrahedra. 
Recently, \citet{xian2019scalable} propose a ``mesh-free'' alternative for tetrahedral meshes. They propose to use farthest point sampling to get ``meshes'' at coarser levels, and define the prolongation using
piecewise constant interpolation which only requires the closest point query. 

\begin{figure}
  \centering
  \includegraphics[width=3.33in]{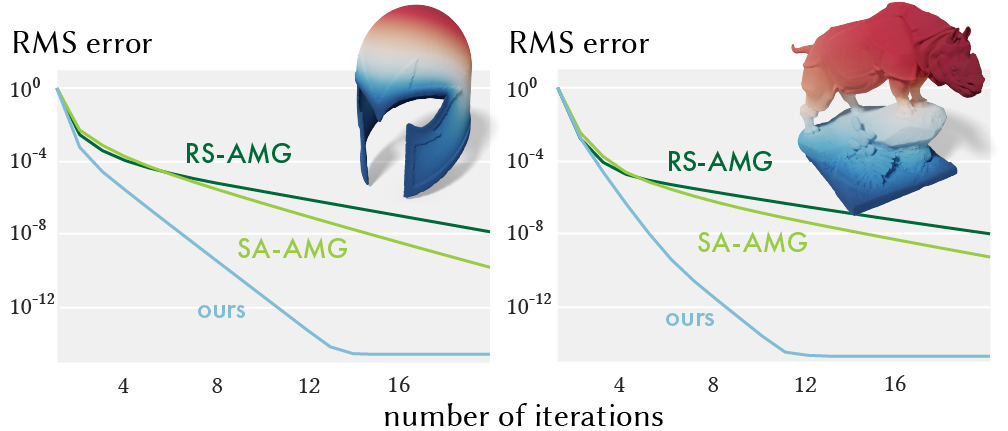}
  \caption{When solving a Poisson problem defined on the surface mesh (top right), we demonstrate that our multigrid based on the intrinsic prolongation (blue) converges faster than the algebraic multigrid methods (green), including the classic Ruge-St\"{u}ben (RS-AMG) \cite{ruge1987algebraic} and the Smoothed Aggregation algebraic multigrid (SA-AMG) \cite{VanekMB96}. Note that we use an off-the-shelf implementation from PyAMG \cite{OlSc2018} with their default multigrid hyperparameters. \textcopyright models by 3DWP (right) under CC BY-SA.}
  \label{fig:amg_compare}
  \vspace{-5pt}
\end{figure}

\paragraph{Algebraic Multigrid}
A popular alternative to deal with unstructured meshes is to use \emph{algebraic
multigrid} \cite{brandt1985algebraic} which builds the hierarchy by treating the
linear system matrix as a weighted graph and coarsening it. This
approach makes no \update{assumptions} on the structure of the geometry. Thus, it is
directly applicable to any unstructured domain. For this reason, algebraic methods are deployed to mesh deformation \cite{ShiYBF06}, cloth
simulation \cite{TamstorfJM15}, and other graphics applications
\cite{KrishnanFS13}. 
However, the cost of algebraic multigrid's generality is the need to re-build
the hierarchy whenever the system matrix changes (see \reffig{fourSteps}).
Furthermore, defining the inter-grid transfer operators for algebraic methods is more challenging and leads to worse performance compared to our method (see \reffig{amg_compare}).

\begin{figure}
  \centering
  \includegraphics[width=3.33in]{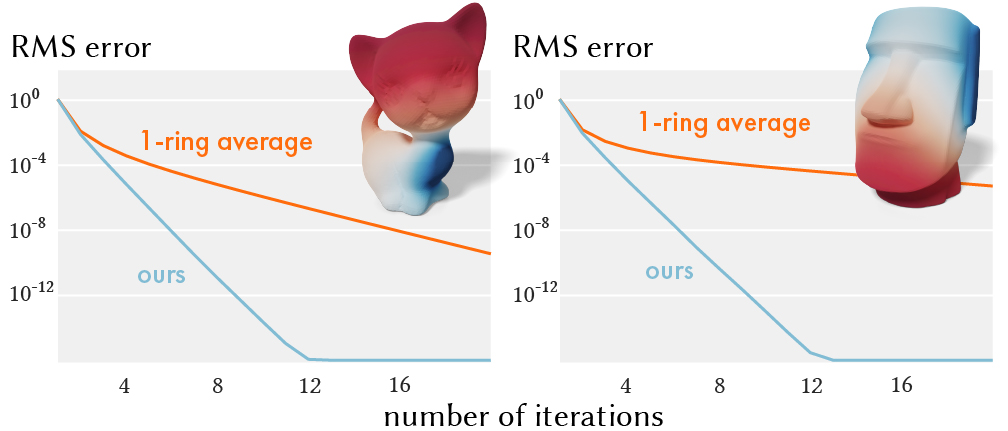}
  \caption{Compared to the prolongation based on the vertex 1-ring average \cite{AksoyluKS05} (orange), our prolongation (blue) leads to a faster convergence rate when solving a Poisson problem on the surface (top right).}
  \label{fig:avg_compare}
  \vspace{-5pt} 
\end{figure} 
\paragraph{Curved Surfaces}
When it comes to surface meshes, defining the prolongation operator becomes more
challenging compared to the Euclidean case. This is because the vertices of a
high-resolution surface mesh do not lie on its coarsened counterpart, thus the
straightforward barycentric computation is not immediately applicable. 
In the special case of subdivision surfaces where the hierarchy is given, there
exists efficient geometric multigrid \cite{GreenTS02} and multilevel
\cite{GoesDMD16} solvers that leverage \update{the} subdivision's regular refinement process
to define \update{the} prolongation.
For unstructured surface meshes, \citet{RayL03} build the hierarchy based on the
progressive meshes \cite{Hoppe96,KobbeltCS98} and define the prolongation
operator \update{using} global texture coordinates. \citet{NiGH04} (and similarly
\citet{ShiBZ09}) find a maximum independent set of vertices to build a hierarchy and
compute a prolongation operator based on \update{a} weighted average among one-ring
neighboring vertices.
\citet{AksoyluKS05} propose several methods for hierarchy construction
based on removing the maximum independent set of vertices. Similarly, they also
compute the prolongation by averaging among the one-ring neighbors. 
These approaches either need additional information (e.g., having subdivision
connectivity or texture information) or use  one-ring average (combinatorial
information) to define the prolongation. But one-ring average often leads to a
denser system because it requires on average 6 vertices to interpolate the
result on a vertex, in contrast to 3 when using the barycentric interpolation.
Performance wise, in \reffig{avg_compare} we show that our prolongation leads to
a better convergence compared to the multigrid based on averaging one-ring
vertices. 

Our method computes \update{the} prolongation based on \emph{intrinsic geometry}, in a similar spirit to \cite{SharpSC19a}. This enables us to define a linear prolongation simply using the barycentric coordinates, echoing the success of using barycentric coordinates in the Euclidean \update{case}. 
Furthermore, our approach allows one to plug-and-play different decimation
strategies to construct multigrid hierarchies (see \reffig{diff_decimation}). This
flexibility allows one to pick a well-suited decimation method for their tasks
of interest. 

Purely algebraic direct solvers (e.g., sparse Cholesky) are the \emph{de
facto} standard in geometry processing due to their reliability, scalability (as memory allows), and precision.
Factorizations can be reused for changing right-hand sides (see
\reffig{fourSteps}), but trouble arises for the myriad applications where the
system matrix also changes.
Special classes of sparse changes can be efficiently executed:
low-rank updates \cite{Chen2008,Cheshmi2020} or partial re-factorizations \cite{Herholz2018cup,HerholzS20}.
However, \update{many other scenarios trigger full numerical
refactorization, such as changing parameters in a multiobjective optimization and Hessian updates for Newton's method}
Due to the overwhelming popularity of Cholesky factorization, we focus on it for many of our head-to-head comparisons.

%% file: sections/multigrid.tex
\section{Multigrid Overview}\label{sec:surfMultigrid}
Multigrid is a type of iterative solver that is scalable to solve large linear systems $\mA \vx = \vb$. In this paper, we propose a novel geometric multigrid method for solving linear systems defined on curved surfaces, represented as irregular triangle meshes. We refer readers to excellent resources on multigrid \cite{trottenberg2000multigrid, brandt2011multigrid}, here we give only the essential components needed to understand our method. Note that we use ``grid'' or ``mesh'' interchangeably to denote the underlying geometry.

\begin{figure}
  \centering
  \includegraphics[width=3.33in]{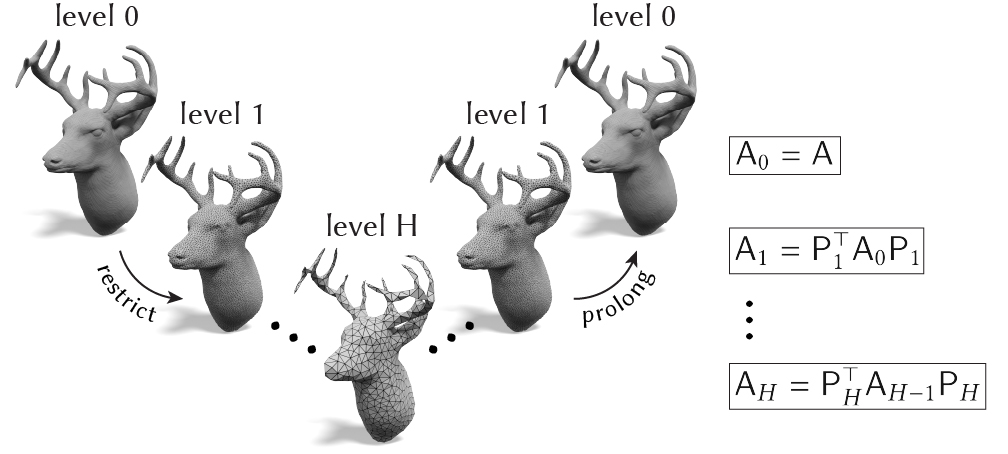}
  \caption{The multigrid V-cycle proceeds from the finest grid (level 0) to the coarsest grid (level H) and \update{goes} back up to the finest grid again. On each level (except \update{for} the coarsest level), we pre-relax the solution, \emph{restrict} it to the coarser grid, compute the correction, \emph{prolong} the correction back to the finer level, post-relax the correction, and then add the correction to the current solution. Our approach belongs to the \emph{Galerkin multigrid} where we define the system matrix at a coarser level as $\mA_h = \mP^\top_{h} \mA_{h-1} \mP_h$. \textcopyright model by Takeshi Murata under CC BY-SA.}
  \label{fig:vcycle}
  \vspace{-5pt}
\end{figure} 
Multigrid methods solve a linear system in a hierarchical manner by employing
two complementary processes: \emph{relaxation} and \emph{coarse-grid
correction}. Relaxation involves applying classic iterative methods to correct
the high-frequency error between the current solution and the exact solution of
the system. Coarse-grid correction involves transferring the low-frequency error
to a coarser mesh through \emph{restriction}, solving a coarse-grid system of
equations, then transferring the \update{correction} back to the finer mesh via
\emph{prolongation} (a.k.a. \emph{interpolation}). This process of going from
the fine grid to the coarse grid and then back to the fine grid is called the
\emph{V-cycle} (see \reffig{vcycle}). How to build the multigrid hierarchy and
how to transfer information back and forth between grid levels are \update{keys to determine} the efficiency of a multigrid method.

\begin{figure*}
  \centering
  \includegraphics[width=7.0in]{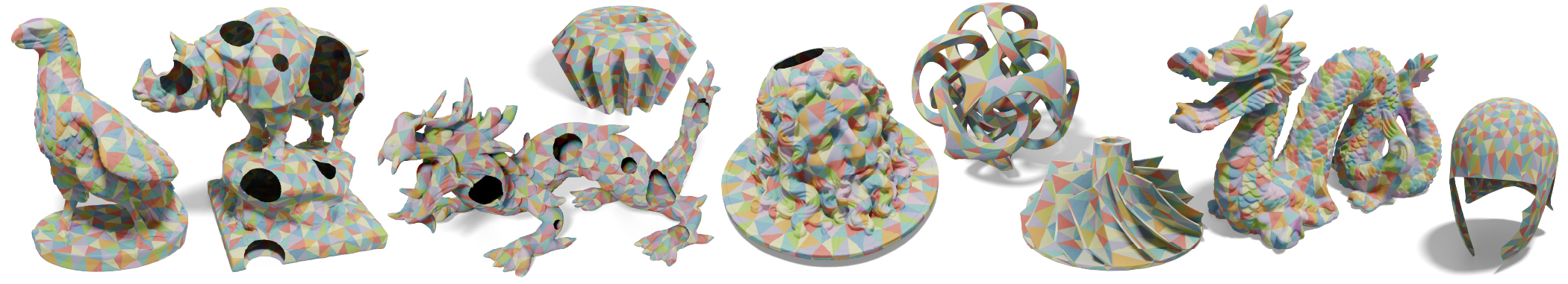}
  \caption{We visualize the bijective map computed using our method by coloring the high-resolution shape using the coarsened triangulation (as different colors). Our method is applicable to man-made objects, organic shapes, high-genus shapes, and meshes with boundaries. \textcopyright models by Oliver Laric (left 1, 5, 8) under CC BY-NC-SA and Landru (left 7) under CC BY.}
  \label{fig:SSP_visualization}
  \vspace{-5pt}
\end{figure*}
Our method belongs to \emph{geometric multigrid} based on the \emph{Galerkin coarse grid approximation}. Geometric multigrid is a class of multigrid methods that builds the hierarchy purely based on the geometry, requiring no knowledge about the linear system. Galerkin coarse grid approximation builds the system matrix $\mA_c$ on the coarsened mesh from the system matrix $\mA$ on the original mesh as
\begin{align}\label{equ:gca}
  \mA_c = \mR \mA \mP,
\end{align}
where $\mR$ is the restriction operator to transfer signals from the fine mesh
to the coarsened mesh and $\mP$ is the prolongation operator to transfer signals
from coarse to fine. \update{When} $\mA$ is symmetric, \update{many methods often}
define $\mR = \mP^\top$. Thus, defining the prolongation operator $\mP$ is
extremely critical for Galerkin multigrid because it determines both the quality
of the coarsened linear system $\mP^\top \mA \mP$ and the quality of the
inter-grid transfer (restriction $\mP^\top$ and prolongation $\mP$). 
In \refalg{vcycle}, we provide pseudo code of the Galerkin multigrid V-cycle
where $\mP$ plays a crucial role in the entire algorithm.
An ideal prolongation must accurately interpolate smooth functions (low distortion) to ensure fast convergence. The prolongation also needs to be sparse to enhance the solver efficiency at coarser levels. 

Defining a prolongation that satisfies these properties is well-studied on structured domains, but extending to unstructured curved surfaces \update{remain} a challenging problem until now. 
In this paper, we use successive self-parameterization to compute a prolongation
operator $\mP$ for curved surfaces based on the intrinsic geometry. Our novel
joint flattening further reduces the distortion caused by the parameterization
and our extension to \update{meshes} with boundaries broadens the applicability of our
method to many complex geometries. When deploying our prolongation to the
Galerkin multigrid framework, our method achieves \update{better} convergence
over alternative multigrid methods for curved surfaces.
%


%% file: sections/prolongation.tex
\section{Intrinsic Prolongation}
\begin{figure}
  \centering
  \includegraphics[width=3.33in]{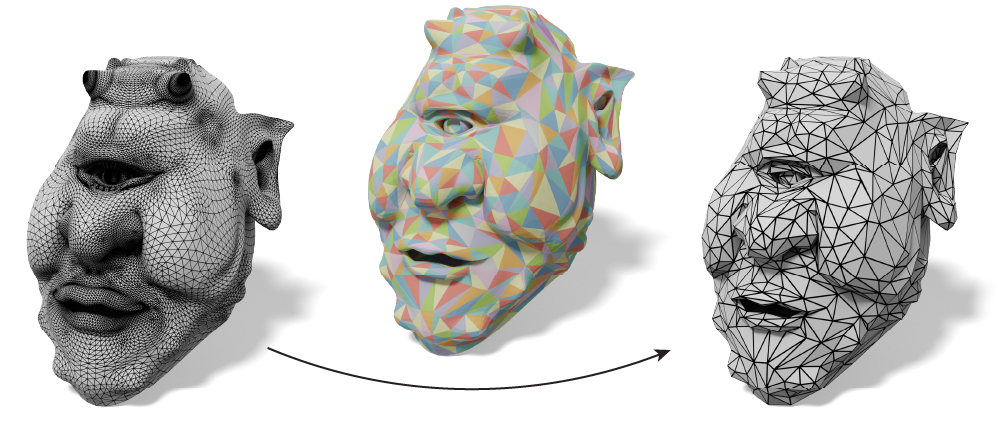}
  \caption{Given a high-resolution shape (left) and its coarsened counterpart (right), we compute a bijective map between the two so that for any given point on the fine mesh we can compute its corresponding barycentric coordinates on the coarsened mesh, and vice versa. We visualize the map by coloring the coarse triangulation on top of the high-resolution model. }
  \label{fig:bijectiveMap}
  \vspace{-5pt}
\end{figure}

The central ingredient of Galerkin multigrid is the prolongation operator to
interpolate signals from a coarsened mesh to its fine version. 
We compute the prolongation by maintaining an intrinsic
parametrization, as opposed to extrinsic prolongtaion based on 3D spatial
coordinates (cf.~\cite{MansonS11a,CubicStyle2019}).
Specifically, we parameterize the high-resolution mesh using the coarsened mesh to obtain a bijective map between the two (see \reffig{bijectiveMap}). Given a point on the high-resolution mesh, we can obtain its corresponding barycentric coordinates on the low-resolution mesh, and vice versa. We can then assemble a linear prolongation operator based on the barycentric information. 

We compute the bijective map using \emph{successive self-parameterization}. The key idea is to successively build a bijective map for each decimation step and assemble the full map via compositing all the maps. 
Our method for computing the successive parameterization is based on the framework of \cite{LiuKCAJ20}, which can be perceived as a \update{combination of \citet{LeeSSCD98} and \citet{CohenMO03}}. The key differences of our method compared to \cite{LiuKCAJ20} are a novel \emph{joint flattening} method (see \refsec{jointFlattening}) to further reduce the distortion and a generalization to meshes with boundaries (see \refsec{boundary}). For the sake of reproducibility, we reiterate the main ideas of successive self-parameterization here.

\subsection{Successive Self-Parameterization}
Let $\M^0$ be the input fine mesh with/without boundary, and $\M^0$ is successively simplified into a series of meshes $\M^l$ with $0 ≤ l ≤ L$ until the coarsest mesh $\M^L$. For each pair of meshes $\M^l, \M^{l+1}$, we use $f^{l}_{l+1}: \M^{l} → \M^{l+1}$ to denote the bijective map between them. 
The main idea is to compute each $f^{l}_{l+1}$ on-the-fly during the decimation process and composite all the maps between subsequent levels to obtain the final map $f^0_L: \M^0 → \M^L$ as
\begin{align} \label{equ:composition}
  f^0_L =   f^{L}_{L+1} ∘ \cdots ∘ f^0_1.
\end{align}
Thus, the question boils down to the computation of the individual maps $f^{l}_{l+1}$ before and after a single edge collapse.

\begin{figure}
  \centering
  \includegraphics[width=3.33in]{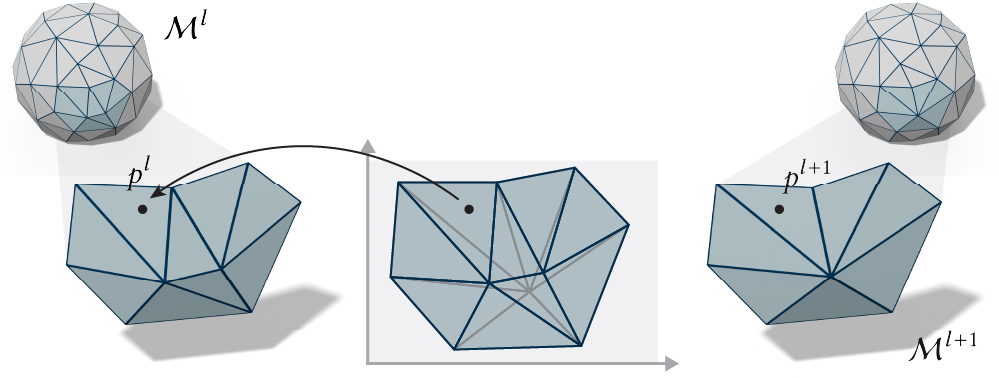}
  \caption{Since both the edge 1-ring before the collapse (left) and the vertex 1-ring after the collapse (right) are mapped to the same 2D domain with a consistent boundary curve (middle), we can easily use the shared UV space to map a point back and forth between $\M^l$ and $\M^{l+1}$.}
  \label{fig:baryQuery}
  \vspace{-5pt}
\end{figure}

For each edge collapse, the triangulation mostly remains the same except for the neighborhood of the collapsed edge. Thus, computing $f^{l}_{l+1}$ \update{only requires to figure} out the mapping within the edge 1-ring neighborhood. Let $\N^l(k)$ be the neighboring vertices of a vertex $k$ (including vertex $k$ itself) at level $l$ and let $\N^l(i,j) = \N^l(i) \cup \N^l(j)$ denote the neighboring vertices of an edge $i,j$. 
The key observation is that the boundary vertices of $\N^l(i,j)$ before the collapse are the same as the boundary vertices of $\N^{l+1}(k)$ after the collapse, where $k$ is the newly inserted vertex after collapsing edge $i,j$.
%
%
Hence, we compute a shared UV-parameterization for the patches enclosed by $\N^{l}(i,j)$ and $\N^{l+1}(k)$ with the same boundary curve. Then, for any given point $p^{l} ∈ \M^{l}$ (represented in barycentric coordinates), we can utilize the shared UV parameterization to map $p^{l}$ to its corresponding barycentric coordinates $p^{l+1} ∈ \M^{l+1}$ and vice-versa, as shown in \reffig{baryQuery}.

\subsection{Joint Flattening}\label{sec:jointFlattening}
The base method proposed by \citet{LiuKCAJ20} ensures boundary consistency by first flattening the edge 1-ring $\N^l(i,j)$ and \update{setting} the boundary vertices \update{as hard constraints} when flattening the vertex 1-ring $\N^{l+1}(k)$ after the collapse. Although this method can 
\begin{wrapfigure}[15]{r}{0.4\linewidth}
  \vspace*{-0.1\intextsep}
  \hspace*{-0.7\columnsep}
  \begin{minipage}[b]{1.15\linewidth}
  \includegraphics[width=0.9\linewidth, trim={0mm 4mm 0mm 0mm}]{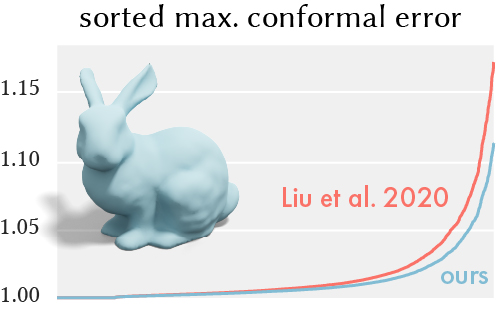}
  \caption{For each edge 1-ring on this bunny mesh, we collapse the edge and flatten the patch using the method of \cite{LiuKCAJ20} and our joint flattening method. We visualize the sorted quasiconformal distortion among all the 1-rings and demonstrate that our method (blue) leads to less distortion.}
  \label{fig:jointFlat}
  \end{minipage}
\end{wrapfigure}
ensure boundary consistency, it always favors minimizing the distortion of $\N^l(i,j)$ and creates larger distortion when flattening $\N^{l+1}(k)$. 

We instead compute the shared UV-parameterization by \emph{jointly} minimizing \update{a} distortion energy $E$ defined on the edge 1-ring $\N^l(i,j)$ before the collapse and the vertex 1-ring $\N^{l+1}(k)$ after the collapse while ensuring boundary consistency.
\update{In \reffig{jointFlat}, we demonstrate that our joint flattening results in a parameterization with less distortion compared to the method by \citet{LiuKCAJ20}.}

For the notational convenience, we use $\Vb, \Fb$ to denote the vertices and faces of the local patch within $\N^l(i,j)$ before the collapse, and $\Va, \Fa$ to denote the vertices and faces of the local patch within $\N^{l+1}(k)$ after the collapse.
We then write the joint energy optimization problem as
\begin{align}\label{equ:jointEnergy1}
  &\minimize_{\U^l, \U^{l+1}}\ E(\Vb, \Fb, \U^l) + E(\Va, \Fa, \U^{l+1}) \\
  &\text{ subject to } \ub_b = \ua_b
\end{align}
where we use $\U^l ∈ \R^{|\Vb|×2}$ to represent the UV locations of $\Vb$ at level $l$ and each $\vu_i^l ∈ \R^2$ denotes a UV-vertex position. We also use $\ub_b, \ua_b$ to represent the boundary vertices of $\N^l(i,j)$ and $\N^{l+1}(k)$, respectively.

\begin{figure}
  \centering
  \includegraphics[width=3.33in]{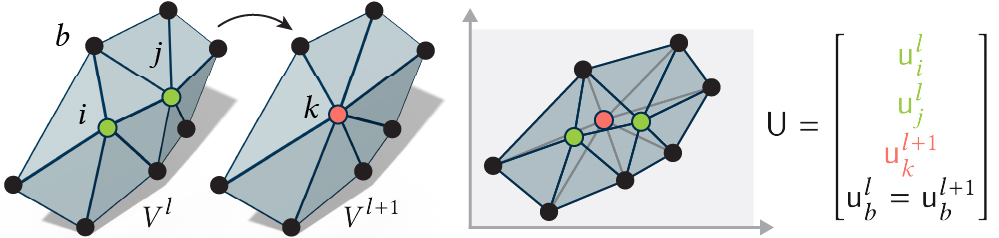}
  \caption{Ensuring bijectivity between $\Vb$ and $\Va$ requires their boundary vertices (black $b$) to have the same UV positions (right). We handle this constraint by introducing a joint variable $\U$ which \update{contains} shared degrees of freedom on the boundary.}
  \label{fig:jointVar}
  \vspace{-5pt}
\end{figure}

In order to handle the constraints, we introduce a joint variable $\U = \U^l ∪ \U^{l+1}$ (see \reffig{jointVar}) to incorporate the equalities into the degrees of freedom and turn \refequ{jointEnergy1} into an unconstrained problem 
%
%
\begin{align}\label{equ:jointEnergy}
  &\min_{\U}\ \ E(\Vb, \Fb, \U) + E(\Va, \Fa, \U)
\end{align}    
Introducing the joint variable $\U$ allows us to minimize the distortion energy for the patch before and after the collapse simultaneously. 
%

\subsection{Boundary Edges}\label{sec:boundary}
The creation of the joint variable in \reffig{jointVar} is only applicable when
the collapsed edge lies fully in the interior of \update{the} triangle mesh. 
For boundary edges,
different treatment is required to ensure the shared parameterization has a
consistent boundary curve in order to preserve the bijectivity (cf.~\cite{LiuFerguson2017}).

\begin{figure}
  \centering
  \includegraphics[width=3.33in]{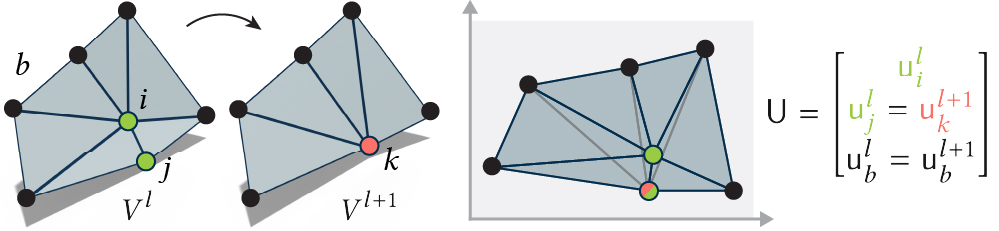}
  \caption{When one of the edge vertices is on the boundary (vertex $j$ in this case), we have to also constrain the vertex $j$ and $k$ to have the same UV location to ensure bijectivity, as shown in the joint variable $\U$ on the right.}
  \label{fig:jointVar_case1}
  \vspace{-5pt}
\end{figure}

In the case where one of the two incident vertices lies on the boundary, we create a
joint variable which snaps the UV position of the other (interior) vertex to the boundary (see \reffig{jointVar_case1}).
Note that we only perform this snapping operation in
the parameterization domain, their corresponding vertices $\vv_j, \vv_k$ in
$\R^3$ are still placed at the locations which minimize the decimation error
metric (e.g., appearance preservation).
\begin{figure}
  \centering
  \includegraphics[width=3.33in]{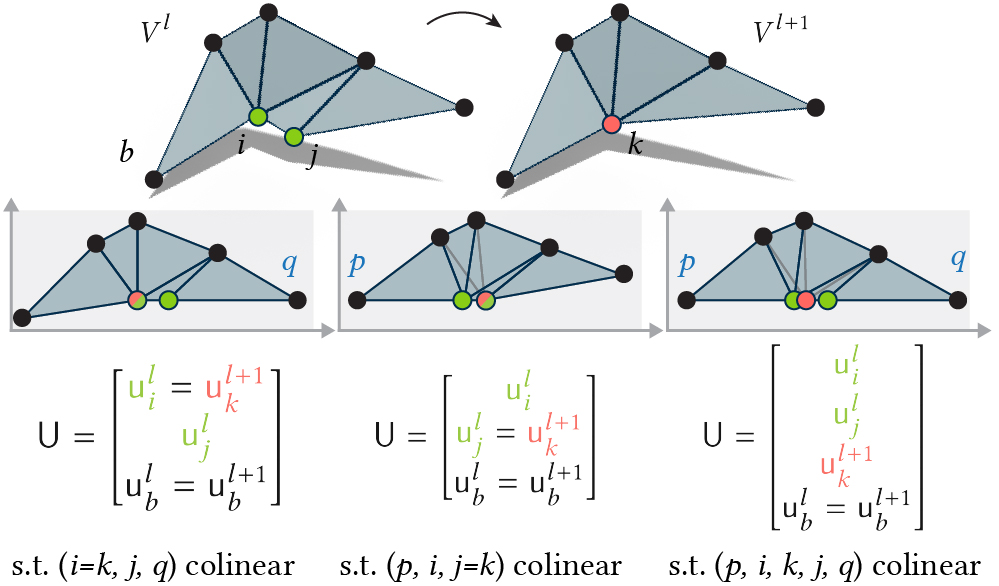}
  \caption{When the edge $i,j$ is a boundary edge, we consider three cases: $\ub_i = \ua_k$ (left), $\ub_j = \ua_k$ (middle), and vertex $i,j,k$ are colinear in the UV space (right). To ensure the boundary curves remain consistent, these cases result in three different sets of colinearity constraints (see the bottom row), where we use $q$ to represent the \emph{next} boundary vertex of the edge $i,j$ and we use $p$ to represent the \emph{previous} boundary vertex of the edge $i,j$. }
  \label{fig:jointVar_case2}
  \vspace{-5pt}
\end{figure} 

In the case where both edge vertices are on the boundary, we determine the joint
variable $\U$ by choosing best of three possible choices.
\update{Suppose} the boundary edge $i,j$ is collapsed to a boundary vertex $k$, we consider
the cases where (1) vertex $k$ lies on vertex $i$, (2) vertex $k$ lies on vertex
$j$, and (3) vertex $k$ lies on the line defined by $i,j$. Even though case (1),
(2) seem unnecessary when we have case (3), these cases end up with different
sets of constraints in the joint flattening optimization. Thus, we consider all
three cases and take the one with the minimum energy value. In
\reffig{jointVar_case2}, we show how we group variables for the three cases and
their corresponding colinearity constraints to maintain the same boundary curve.
We impose the colinearity via adding Dirichlet constraints $(\vu_i)_y = 0$ for
all the vertices $i$ that are
colinear. 
%

\subsection{Decimation Strategies \& Distortion Energies}\label{sec:generalization}
Our joint flattening makes no assumption on the edge collapse algorithm in use.
For instance, one could use the \emph{quadric error edge collapse} (qslim)
\cite{GarlandH97} to preserve the appearance, \emph{mid-point edge collapse} to
encourage the coarse triangulation to be more uniformly distributed, and the
\emph{vertex removal} (via half-edge collapses \cite{KobbeltCS98}) to ensure \update{that the} vertices on the coarsened mesh \update{are} a subset of the fine vertices. 

\begin{figure}
  \centering
  \includegraphics[width=3.33in]{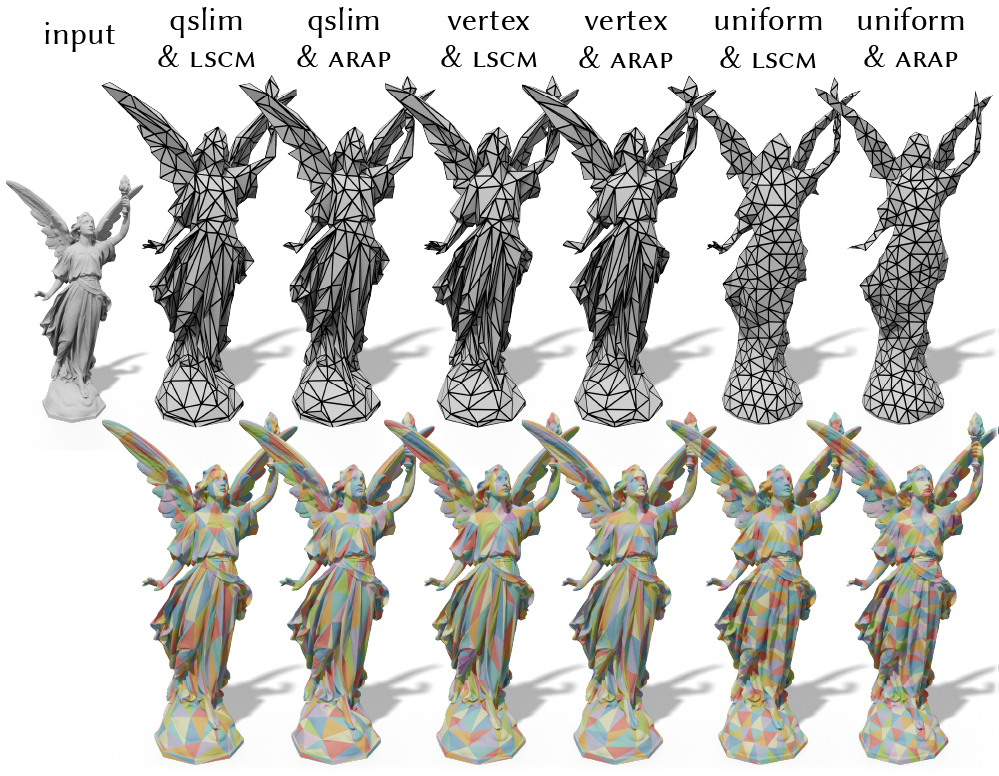}
  \caption{Our method allows to plug-and-play different decimation strategies and parameterization algorithms. In these examples, we decimate the model using qslim \cite{GarlandH97}, vertex removal via half-edge collapse, and uniform mid-point edge collapse. We use \arap \cite{LiuZXGG08} and \lscm \cite{LevyPRM02} as the parameterization algorithms. The influence of these combinations to the solver convergence is shown in \reffig{mg_decType}. } 
  \label{fig:diff_decimation}
  \vspace{-5pt}
\end{figure}

The distortion energy $E$ in \refequ{jointEnergy} provides another design
parameter. In \reffig{diff_decimation}, we demonstrate the flexibility of our
joint flattening by minimizing the \emph{as-rigid-as-possible} (\arap)
\cite{LiuZXGG08} and the \emph{least square conformal map} (\lscm)
\cite{LevyPRM02} energies.

\begin{figure}
  \centering
  \includegraphics[width=3.33in]{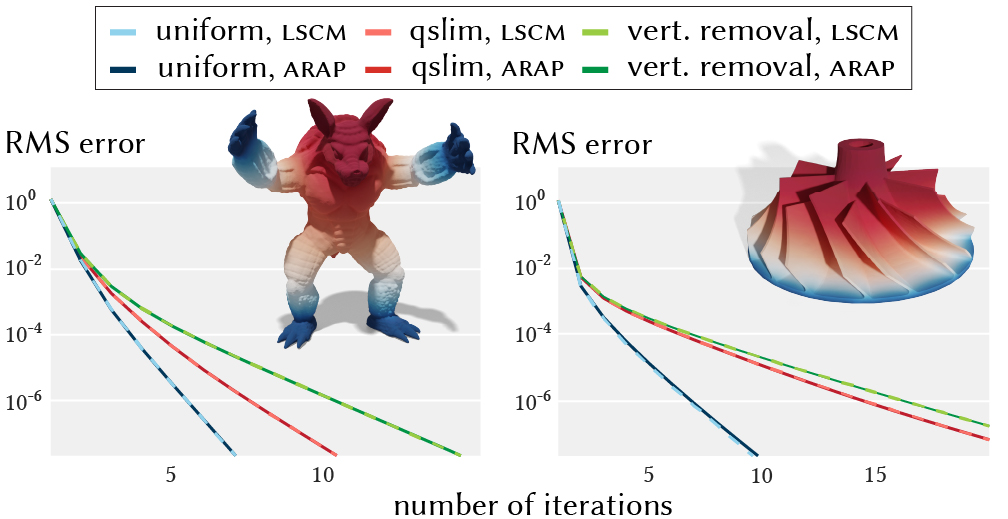}
  \caption{Different combinations of the decimation and the parameterization methods lead to different performance in down-stream applications. For example, in the context of multigrid solvers on a Poisson problem, the uniform edge decimation with \lscm leads to a better convergence rate. }
  \label{fig:mg_decType}
  \vspace{-5pt}
\end{figure}

Depending on the intended application, different combinations of the decimation strategy and the parameterization algorithm may lead to different performance. For instance, in \reffig{mg_decType} we compare the convergence behavior of our Galerkin multigrid solvers constructed using these combinations. In our experiments, using the uniform decimation with \lscm \update{leads} to the best performance among these options. 
Other options for minimizing the distortion \cite{KhodakovskyLS03},
computing the map \cite{GuskovVSS00,GuskovKSS02, FriedelSK04}, \update{and}
decimation strategies (e.g., \cite{TrettnerK20}) seem attractive to combine with
our joint flattening. It is however more challenging and thus left as a future
work.

\subsection{Prolongation Operator}
The above discussion computes a bijective map between a pair of meshes that
undergoes a single edge collapse. We can easily extend the method to compute a
map between two consecutive multigrid levels via composition (see
\refequ{composition}). Given this information, we can now compute a prolongation
operator for surface multigrid.

We choose linear interpolation as our prolongation operator because it is sufficient \update{for the} convergence of the second-order PDEs \update{typically employed in computer graphics} \cite{hemker1990order}. Although some of our experiments consider higher order PDEs, many of them are reduced to low-order systems in practice via mixed finite elements \cite{JacobsonTSZ10}. Empirically, we find that linear prolongation \update{still converges} in most cases. 

\begin{figure}
  \centering
  \includegraphics[width=3.33in]{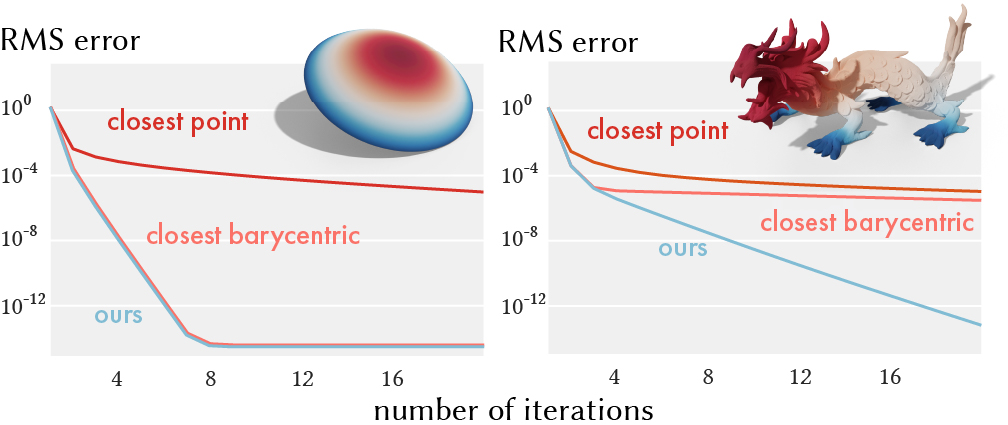}
  \caption{We compare our intrinsic prolongation with naive extrinsic prolongations based on the closest-point projection. When evaluating on a simple shape (left), most methods can converge; when evaluating on a complex shape (right), only our intrinsic prolongation converges. \textcopyright model by Oliver Laric (right) under CC BY-NC-SA.}
  \label{fig:prolong_compare}
  \vspace{-5pt}
\end{figure}

\begin{figure}
  \centering
  \includegraphics[width=3.33in]{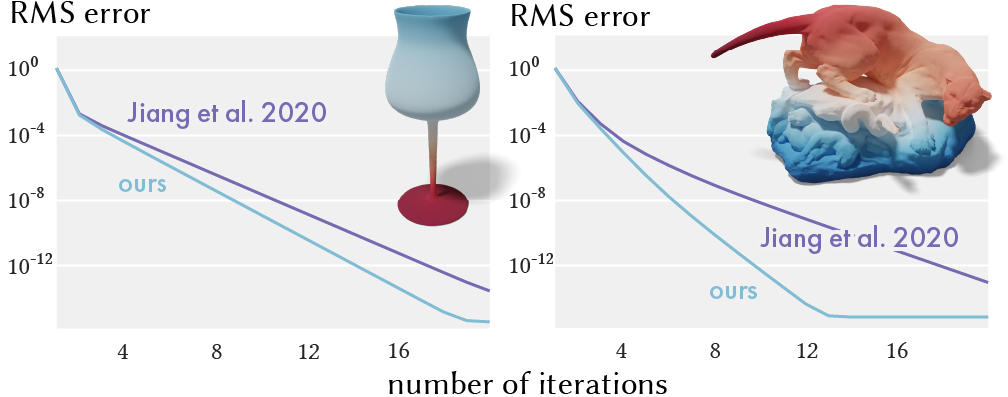}
  \caption{Compared to an extrinsic prolongation \update{proposed by} \citet{JiangSZP20}, our prolongation leads to faster convergence.}
  \label{fig:breve_compare}
  \vspace{-5pt} 
\end{figure}

Our linear prolongation $\mP$ is a tall matrix whose size is the number of fine
vertices by the number of coarse vertices. Each row of $\mP$ contains 3
non-zeros corresponding to the barycentric coordinates of the fine vertex 
with respect to the vertices of the coarse triangle containing it.
We evaluate the quality of our prolongation on solving Poisson problems on a
variety meshes. We demonstrate that our intrinsic prolongation leads to
faster convergence compared to the naive closest point projection
(\reffig{prolong_compare}), an extrinsic bijective projection by
\citet{JiangSZP20} (\reffig{breve_compare}), vertex 1-ring average
\cite{AksoyluKS05} (\reffig{avg_compare}), and algebraic multigrid prolongations
(\reffig{amg_compare}).

%% file: sections/implementation.tex
\begin{figure}
  \centering
  \includegraphics[width=3.33in]{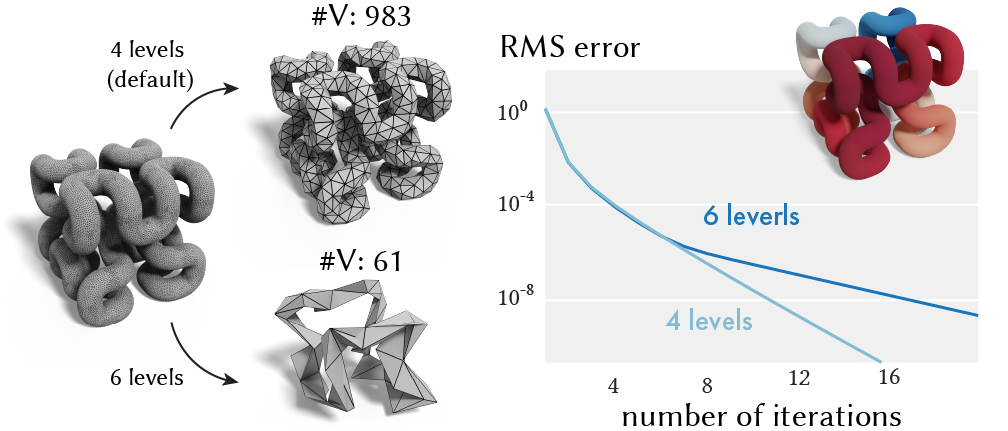}
  \caption{\update{We compare the multigrid convergence between our default parameters (resulting in four multigrid levels) and an extreme coarsening (6 levels). We demonstrate that the extreme coarsening hurts the performance (right) because the coarsest mesh fails to represent the target solution. }}
  \label{fig:coarse_quality}
  \vspace{-5pt}
\end{figure} 
\section{Multigrid Implementation}\label{sec:implementation}
Switching from direct solvers to our multigrid method is very simple. In \refalg{surfaceMultigrid} we summarize how one can implement a Galerkin geometric multigrid method to solve a linear system. The key difference is that the pre-computation happens right after loading the input mesh. This pre-computation takes the vertex and face lists as the inputs, and outputs a series of prolongation operators $\mP_1, \cdots, \mP_H$ for different levels \update{on} the hierarchy. After building the linear system $\mA,\vb$, one can run the V-cycle until getting the desired accuracy.

In \refalg{vcycle}, we summarize the pseudo code of the V-cycle algorithm which consists of two procedures: relaxation and coarse-grid correction. 
For the relaxation step, we use the standard serial Gauss-Seidel method. 
In the coarse-grid correction step, the process is well-defined given the prolongation operator $\mP$. We start by restricting the residual to the coarser level via $\mP^\top$, solving a coarsened linear system with the left-hand-side defined as $\mP^\top \mA \mP$, prolonging the low-res solution back to the fine domain using $\mP$, and using it to update the current high-res solution.
We can further accelerate the computation by storing the system matrix hierarchy $\mA_{h+1} = \mP_{h+1}^\top \mA_h \mP_{h+1}$ to save some redundant computation.

\IncMargin{1em}
\begin{algorithm}[t]
  \SetKwInOut{Input}{Input}
  \SetKwInOut{Output}{Output}
  \SetKwInOut{Parameter}{Param.}
    \caption{Galerkin Surface Multigrid Solver}  
    \label{alg:surfaceMultigrid}
    \Indentp{-1em}
    \Indentp{1em}
      \textit{\color{iglGreen} $\mV,\mF ←$ load triangle mesh}\\
      $\mP_1, \cdots, \mP_H ←$ \textit{precompute multigrid hierarchy $(\mV,\mF)$} \\
      \textit{\color{iglGreen} $\mA ←$ build left-hand-side}\\
      \textit{\color{iglGreen} $\vb ←$ build right-hand-side}\\
      \textit{initialize a solution $\vx$}\\
      \While{error is larger than $\epsilon$}{
        $\vx ← \textit{V-cycle}\ (\mA,\vx,\vb, 0)$ \hfill \text{\color{iglGreen} // see \refalg{vcycle}}\\
      }
\end{algorithm}
\DecMargin{1em} 
\IncMargin{1em}
\begin{algorithm}[t]
  \SetKwInOut{Input}{Input}
  \SetKwInOut{Output}{Output}
  \SetKwInOut{Parameter}{Param.}
    \caption{$\vx_\textit{new} = \textit{V-cycle}\ (\mA,\vx_\textit{old}\ ,\vb, h)$}  
    \label{alg:vcycle}
    \Indentp{-1em}
      \Parameter{
                 $\ \mP_1,\mP_2, \cdots, \mP_H,$ \hfill \text{\color{iglGreen} //  hierarchy of prolongations} \\
                 $\ \mu_\textit{pre},\ \mu_\textit{post}$ \hfill \text{\color{iglGreen} //  pre- and post-relaxation iterations}}
      \Input{$\ \mA,$ \hfill \text{\color{iglGreen} // left-hand-side system matrix} \\ 
             $\ \vx_\textit{old},$ \hfill \text{\color{iglGreen} // current solution} \\ 
             $\ \vb,$ \hfill \text{\color{iglGreen} // right-hand-side of the linear system}\\
             $\ h,$ \hfill \text{\color{iglGreen} // current multigrid level}}
      \Output{$\ \vx_\textit{new}$ \hfill \text{\color{iglGreen} // new solution}}
      \BlankLine
    \Indentp{1em}
      \If {$h$ is not the coarsest level $H$}
      {
        \text{\color{iglGreen} // \textsc{pre-relaxation}}\\
        $\vx_\textit{old}' ← \textit{Relaxation}\ (\mA, \vx_\textit{old}\ , \vb, \mu_\textit{pre})$ \\
        \text{\color{iglGreen} // \textsc{coarse-grid correction}}\\
        $\vr_{h+1} ← \mP_{h+1}^\top (\vb - \mA \vx_\textit{old}')$ \hfill \text{\color{iglGreen} // restrict residual}\\
        $\vc_{h+1}← \textit{V-cycle}\ (\mP_{h+1}^\top \mA\mP_{h+1},\textsf{0},\vr_{h+1}, h+1)$\\
        $\vc_{h}←\mP_{h+1}\vc_{h+1}\ $\hfill \text{\color{iglGreen} // prolong correction} \\
        $\vx_\textit{new}' ←\vx_\textit{old} + \vc_{h}\ $\hfill \text{\color{iglGreen} // update solution}\\
        \text{\color{iglGreen} // \textsc{post-relaxation}}\\
        $\vx_\textit{new} ← \textit{Relaxation}\ (\mA, \vx_\textit{new}'\ , \vb,\mu_\textit{post});$ \\
      }
      \Else{
        $\textit{solve}\ \mA \vx_\textit{new} = \vb$ \hfill \text{\color{iglGreen} // direct solve}\\
      }
      $\textbf{return}\ \vx_\textit{new}$
\end{algorithm}
\DecMargin{1em}
In terms of \update{hyperparameters} of our multigrid method, we conduct an ablation study summarized in \refapp{ablation}. In each V-cycle, we use the Gauss-Seidel relaxation with 2 pre- and post-relaxation iterations. Our default setup coarsens the geometry down to 0.25 of \update{the number of vertices at its previous level} until we reach the coarsest mesh with no lesser than 500 vertices. \update{Note that we do not recommend to coarsen the mesh to an extreme. In \reffig{coarse_quality}, we show that an extremely aggressive coarsening often hurts the performance because the coarsest mesh fails to represent the target solution.}
In terms of stopping criteria, the accuracy $ε$ that allows us to get visually distinguishable results compared to the ground truth depends on the problem and the size of the mesh. In our experiments, we set it to be \update{$10^{-3} ≥ ε ≥ 10^{-5}$}.
Our experiments suggest that the optimal set of parameters that minimizes the wall-clock runtime depends on the geometry and the PDE of interest. But we use our default parameters for all our experiments in \refsec{applications} for consistency.

We implement our algorithm in C++ with Eigen and evaluate our method on a MacBook Pro with an Intel i5 2.3GHz processor. 
In comparison with the Cholesky solver where pre-factorization is required whenever the system matrix $\mA$ is changed, our multigrid solver leads to orders of magnitude speed-ups (see \reffig{chol_compare}).

\paragraph{Boundary conditions}
Our implementation currently supports natural boundary conditions (\reffig{chol_compare} right), zero Neumann boundary conditions (\reffig{chol_compare} left), and the Dirichlet boundary condition. We handle the Dirichlet constrains by reformulating the system using only the unknown variables. This results in a reduced and unconstrained linear system, allowing us to solve it as usual. For more details, please refer to \refapp{Dirichlet}.

\begin{figure}
  \centering
  \includegraphics[width=3.33in]{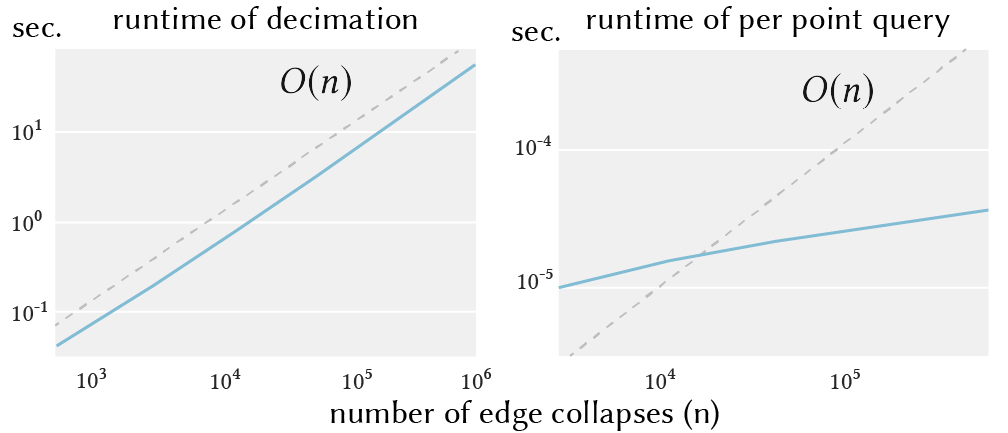}
  \caption{On the left, we report the runtime of our successive \update{self-parameterization} for constructing the bijective map. On the right, we report the query time of a single point through the bijective map.}
  \label{fig:runtime}
  \vspace{-5pt}
\end{figure}

\paragraph{Successive Self-Parameterization}
We report the runtime of our pre-computation (self-parameterization and the query) in \reffig{runtime} and detail the implementation in \refapp{SSP}. Note that this pre-computation solely depends on the geometry. We only need to do this computation once for each shape and we can reuse the same hierarchy for many different linear systems. Thus, in our runtime comparisons in \refsec{applications}, we do not include the runtime of this pre-computation.
\begin{figure}
  \centering
  \includegraphics[width=3.33in]{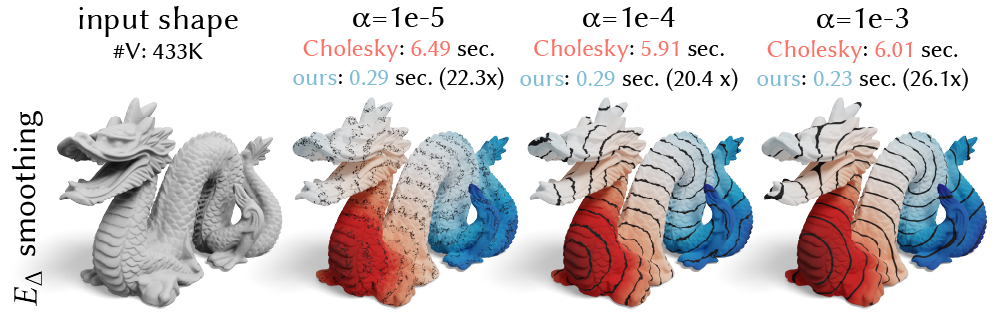}
  \caption{In the data smoothing application, one would adjust the smoothness parameter $α$ until getting the desired smoothness. Using direct solvers (e.g., Cholesky) would need to recompute the expensive pre-factorization, but our multigrid \update{solver} can reuse the same hierarchy and leads to interactive performance. }
  \label{fig:smoothing_tuning}
  \vspace{-5pt}
\end{figure}


%% file: sections/applications.tex
\section{Applications}\label{sec:applications}
We evaluate our method on a variety of geometry processing applications that involve solving linear systems as a subroutine. We especially focus on the case where the system matrix $\mA$ is changing due to different time steps (e.g., simulation) or user interactions (e.g., data smoothing). 
In our experiments, we ignore the multigrid pre-computation and compare our multigrid V-cycle (in blue) against the runtime of both the factorization and the solving time combined of the Cholesky solver (in red) because these steps are required when both $\mA, \vb$ are changing. 
We also pick the applications that involve different system matrices with different sparsity patterns. This includes the cotangent Laplacian (1-ring sparsity), the Bilaplacian (2-ring sparsity), the squared Hessian (2-ring sparsity) \cite{SteinJWG20}, a system matrix derived from Lagrange multipliers \cite{AzencotVWRB15}, and also the Hessian matrices from shell simulation which has $3|V|$-by-$3|V|$ dimensionality. 

\begin{figure}
  \centering
  \includegraphics[width=3.33in]{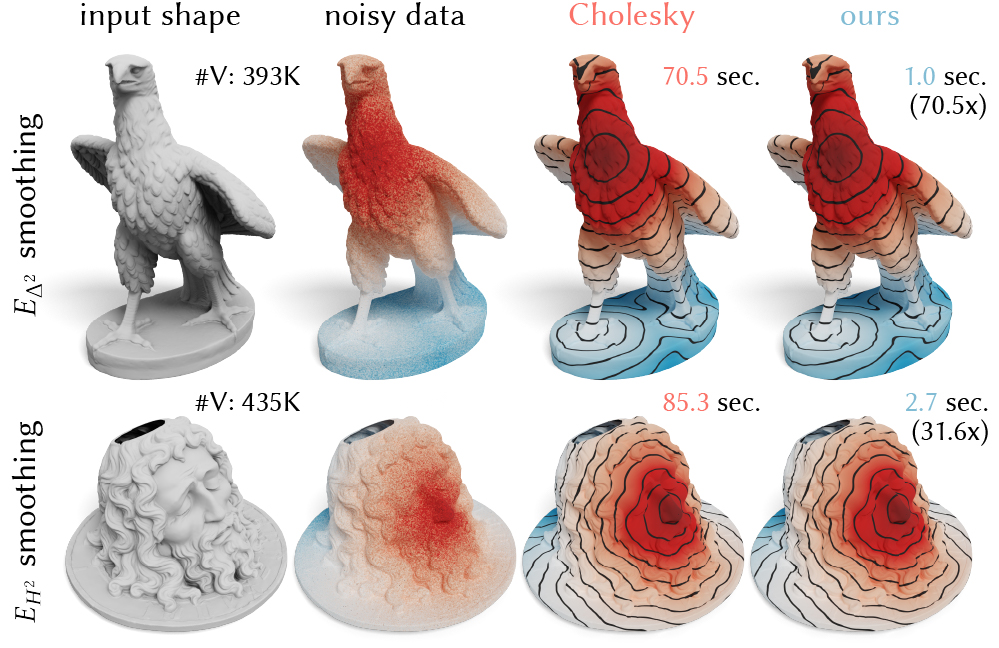}
  \caption{We evaluate our method on data smoothing with different smoothness energies, including the Bilaplacian $E_{Δ^2}$ and the squared Hessian $E_{H^2}$. Our method is orders of magnitude faster than the direct solver. }
  \label{fig:smoothing_preview}
  \vspace{-5pt}
\end{figure} 
\begin{figure}
  \centering
  \includegraphics[width=3.33in]{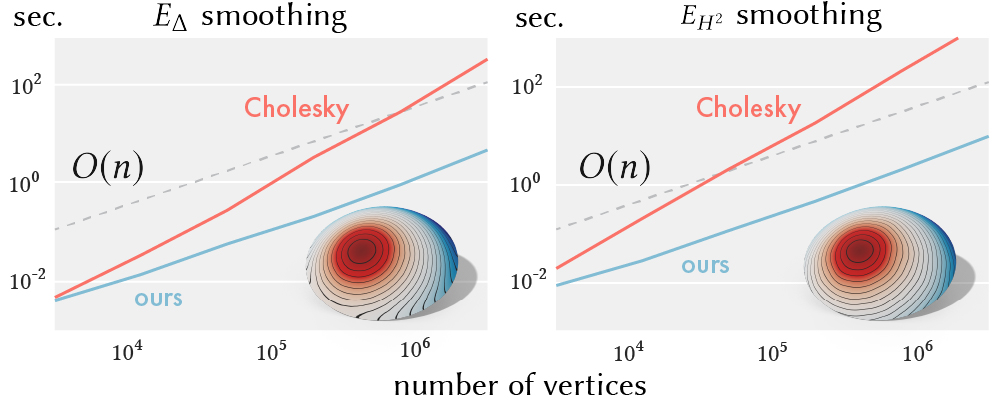}
  \caption{We compare the runtime of our multigrid solver against the Cholesky solver
  on smoothing the noisy data on a sphere cap at different resolutions \update{until reaching a sufficiently small mean squared error (visually indistinguishable)}. We
  evaluate the smoothing with the Dirichlet energy $E_Δ$ (1-ring sparsity) and with the squared Hessian energy $E_{H^2}$ \cite{SteinJWG20} (2-ring sparsity). Our method is asymptotically faster than the direct solver.
  On meshes with 200K vertices and 3 million vertices (using $E_{H^2}$), a serial implementation of our method is 39$\times$ and 231$\times$ faster,
  respectively.}
  \label{fig:chol_compare}
  \vspace{-5pt}
\end{figure} 
\paragraph{Data smoothing} 
Smoothing data on the surface is a fundamental task in geometry processing. We often treat it as an energy minimization problem
\begin{align}\label{equ:dataSmoothing}
  x = \argmin_{x}\ α E_s(x) + (1-α) ∫_Ω \| x - f \|^2 dx,
\end{align}
where $\alpha$ is the parameter controlling the smoothness, $f$ is the input
noisy function, and $E_s$ is an energy of choice, measuring the smoothness of
the output signal $x$. 
As a different input $f$ may contain a different amount of noise, one would
typically adjust the $α$ or the smoothness energy $E_s$ until getting the
desired smoothness. However, these adjustments boil down to solving a different
linear system. 
When using direct solvers, this requires to recompute the factorization in order
to solve the system. In comparison, using our multigrid allows one to reuse the
same precomputed multigrid hierarchy and leads to orders of magnitude speed-ups
(see \reffig{smoothing_tuning}).
%
%
We evaluate our method on different smoothness energies, including the Dirichlet
energy $E_Δ$ (\reffig{smoothing_tuning}), the squared Laplacian energy $E_{Δ^2}$
(\reffig{smoothing_preview} top), and the squared Hessian energy $E_{H^2}$
\cite{SteinJWG20} (\reffig{smoothing_preview} bottom).
%
%
In \reffig{chol_compare}, we quantitatively evaluate the runtime on the same
shape at different resolutions obtained via subdivision. On a mesh with millions of
vertices, our approach has over 100$\times$ speed-ups. 
With our multigrid setup, the precomputed prolongation operator can be reused
not only when changing the value of $\alpha$ (full rank update to $\mA$), but
also when swapping between energies $E_s$.

\begin{figure}
  \centering
  \includegraphics[width=3.33in]{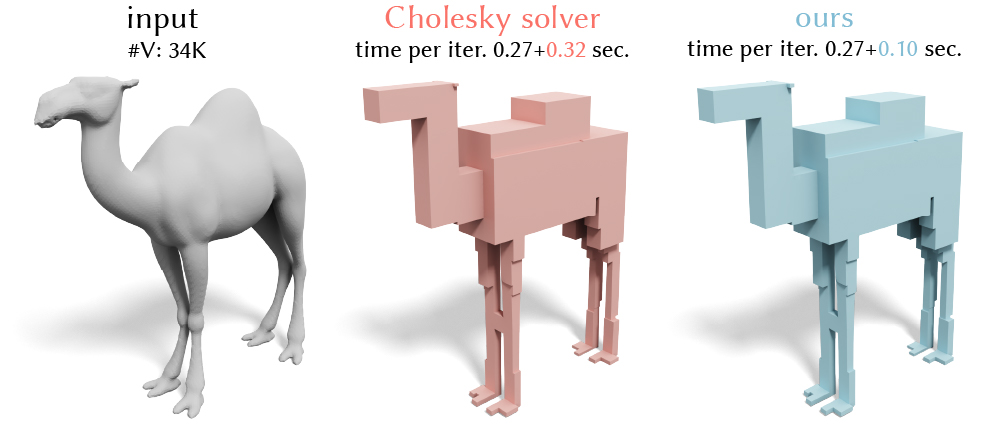}
  \caption{We compare our multigrid against the direct solver on the polycube deformation proposed by \cite{ZhaoLLZXG17}. Although the vertex positions are changed at every iteration, we can still reuse the precomputed multigrid hierarchy because the connectivity remains the same. We report the runtime of other steps in the algorithm in black and the runtime for solving linear systems in red and in blue. }
  \label{fig:polycube}
  \vspace{-5pt}
\end{figure} 
\begin{figure}
  \centering
  \includegraphics[width=3.33in]{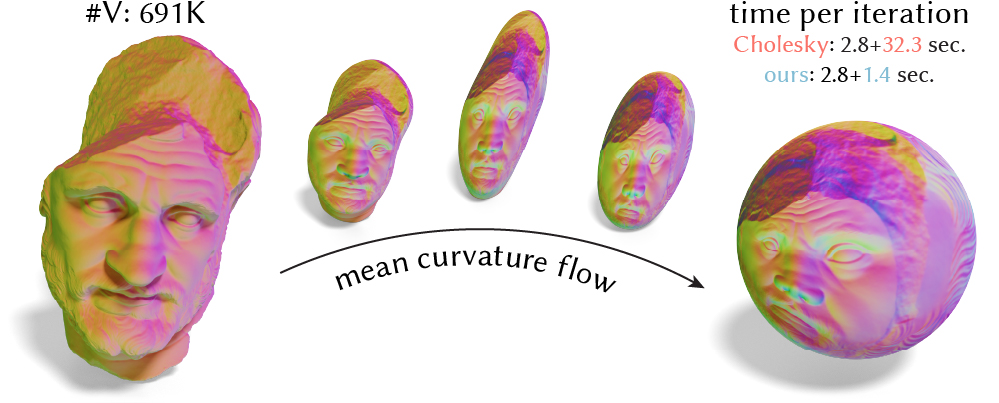}
  \caption{Running \update{the} mean curvature flow \cite{KazhdanSB12} requires to update the system matrix at every step according to the mass matrix of the current mesh. By reusing the hierarchy computed on the input shape (left), our multigrid method is orders of magnitude faster than the direct solver. We report the runtime of other subroutines in black and the time for solving the linear system in red (direct solver) and in blue (our multigrid).  \textcopyright model by Oliver Laric under CC BY-NC-SA.}
  \label{fig:mcf}
  \vspace{-5pt}
\end{figure}
\paragraph{Mesh deformation} 
We also evaluate our method on mesh deformations to demonstrate that even though the vertex positions have changed, as long as the connectivity of the mesh remains the same, we can still reuse the same multigrid hierarchy computed on the rest mesh.
One possible intuition is to view the deformation field on vertices as a function on the rest shape. Thus, a hierarchy built on the rest shape could still be used to compute the deformation ``function''.
In \reffig{polycube}, we evaluate our method on a polycube deformation method proposed by \cite{ZhaoLLZXG17} whose system matrix is \update{re-built} at every iteration based on the current deformed mesh. Our method accelerates the algorithm by 3.2 $\times$ on a relatively low-resolution mesh.
In \reffig{mcf}, we replace the Cholesky solver with our method on a mean curvature flow method proposed in \cite{KazhdanSB12} and achieve 23$\times$ speedup. 

\begin{figure}
  \centering
  \includegraphics[width=3.33in]{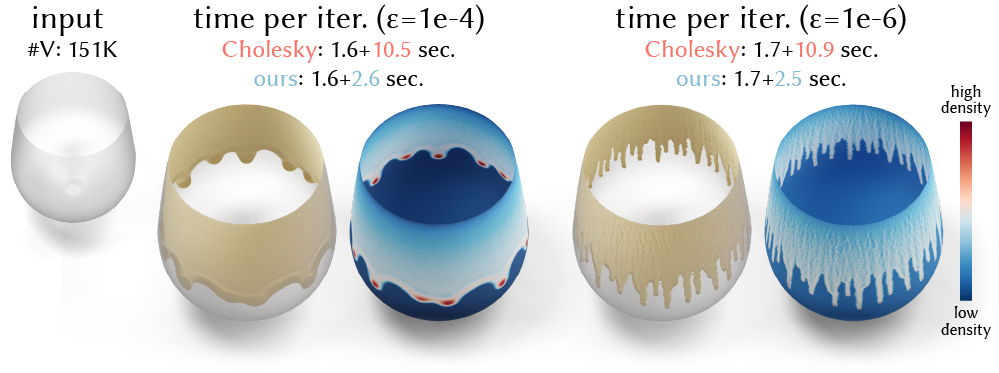}
  \caption{The surface fluid simulation \cite{AzencotVWRB15} involves solving different linear systems at each time step. Our method reuses the precomputed hierarchy and leads to a faster solver in contrast to the direct solver. We split the runtime of other procedures (black) and the runtime of solving the linear system (red and blue). Note that this runtime comparison is in MATLAB using the original implementation from \cite{AzencotVWRB15}. }
  \label{fig:surf_fluid} 
  \vspace{-5pt}
\end{figure}

\begin{figure}
  \centering
  \includegraphics[width=3.33in]{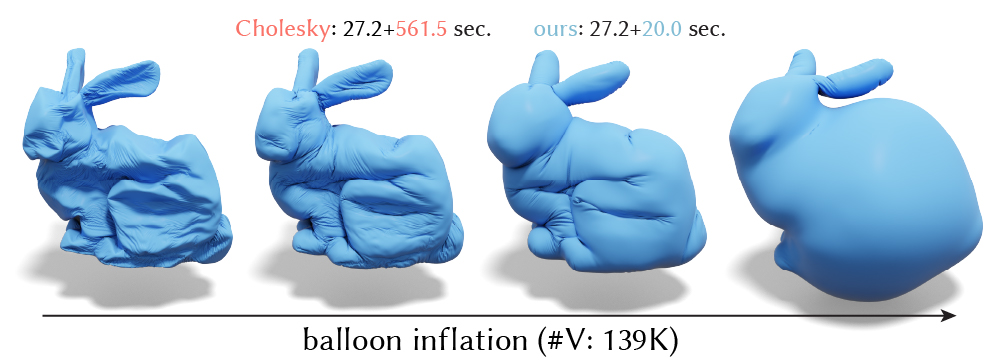}
  \caption{Replacing the Cholesky solver with our surface multigrid method, we can accelerate the linear solve part in the balloon simulation proposed by \cite{SkourasTBG12} by 28$\times$ so that solving linear systems becomes no longer the bottleneck of the algorithm.}
  \label{fig:balloon}
  \vspace{-5pt}
\end{figure}

In many simulation algorithms, the system matrix $\mA$ changes at every time step. 
In \reffig{surf_fluid}, we demonstrate the speed-up of our multigrid solver on a surface fluid simulation \cite{AzencotVWRB15}. Note that the surface fluid simulation is evaluated in \textsc{Matlab} (for both the direct solver and our multigrid) respecting the original implementation.
In \reffig{balloon}, we evaluate our method on a balloon simulation method proposed by \cite{SkourasTBG12}. Due to the speedup of our multigrid method, we shift the bottleneck of balloon simulation away from solving linear systems. 

%% file: sections/discussion.tex
\section{Discussion}
\begin{wraptable}[12]{r}{0.47\linewidth}
    \vspace{-12pt}
    \setlength{\tabcolsep}{5.425pt}
    \centering 
    \caption{Multigrid runtime.$\quad$}
    \vspace{-5pt}
    \hspace*{-10pt}
    \begin{tabularx}{1.06\linewidth}{l|c} 
      \toprule
      \rowcolor{white!50}
      profile (sec.) & \reffig{mcf}  \\
      \midrule
      \rowcolor{derekTableBlue}
      precompute   & 50.6 \\
      total solve time & 1.44 \\
      \rowcolor{derekTableBlue}
      \ 1. prepare $\mP^\top \mA \mP$ & 0.72  \\
      \ 2. relaxation & 0.38 \\
      \rowcolor{derekTableBlue}
      \ 3. prolong \& restrict & 0.10  \\
      \ 4. get residual norm &  0.10 \\
      \rowcolor{derekTableBlue}
      \ 5. others &  0.14 \\
      \bottomrule
    \end{tabularx}
    \label{tab:profile}
\end{wraptable}
In our experiments in \refsec{applications}, we evaluate our runtime using a simple serial implementation of our method. In \reftab{profile}, we further provide a detailed runtime decomposition of the experiment in \reffig{mcf} as a representative example. 
We can observe that preparing the matrix hierarchy $\mP^\top \mA \mP$ and doing the relaxation take most of the time when solving a linear system. Thus, we can achieve even more speedup if we leverage the structure of the problem when computing the matrix hierarchy, such as the data smoothing detailed in \refapp{fastDataSmoothing}, or a parallel implementation of the entire V-cycle.
To validate our hypothesis, our initial attempt uses CPU to parallelize the Gauss-Seidel method based on graph coloring. This reduces the runtime of our relaxation from 0.38 seconds down to 0.17 seconds (2.2$\times$ speedup) for the experiment in \reffig{mcf}. Similarly, for the top and the bottom examples in \reffig{smoothing_preview}, the fast Gauss-Seidel accelerates the relaxation by 1.8$\times$ and 3.3$\times$ repetitively. We provide details about our graph coloring Gauss-Seidel in the \refapp{graphColoring} for completeness. An even higher speed-up can be expected via a GPU implementation of the Gauss-Seidel relaxation (cf.~\cite{FratarcangeliTP16}).
Besides the Gauss-Seidel relaxation, parallelizing the entire solver could also be an interesting future direction to accelerate our method.

%% file: sections/conclusion.tex
\vspace{-5pt}
\section{Limitations \& Future Work}
%
We present a geometric multigrid solver for triangulated curved surfaces. Multigrid methods are asymptotically faster than direct solvers, thus it offers a promising direction for \emph{scalable} geometry processing.
Our multigrid method can obtain a fast approximation of the solution with orders of magnitude speedup. 
\update{However, obtaining a highly accurate solution would require more iterations which results in a less significant speed-ups.
For higher-order problems, our method may not converge to high accuracy because our choice of linear interpolation is insufficient \cite{hemker1990order}. Thus, exploring high-order prolongation (e.g., subdivision barycentric coordinates \cite{anisimov2016subdividing}) or learning-based prolongation (e.g, \cite{KatrutsaDO20}) would also be valuable directions to improve the solver.}
\update{Another interesting direction to improve the solver} is to use our multigrid solver as the pre-conditioner for other solvers such as the \emph{conjugate gradient} method.

%
%
%
Developing a reliable and robust surface multigrid solver would be an important next step.
Our current solver is more sensitive to the triangle quality of the input mesh compared to the existing direct solver. In our experiments, we \update{remesh} troublesome input shapes using available methods \cite{JakobTPS15,HuSWZP20,meshmixer}.
A better future approach would be extending our self-parameterization to the
entire remeshing process, to maintain bijectivity from the remeshed object to
the input mesh. 
Having a deeper understanding of the relationship between the convergence and
mesh quality would give insights towards developing a suitable remeshing
algorithm for surface multigrid solvers. 
Achieving this may also require theoretical tools to estimate the convergence property, such as extending the \emph{Local Fourier Analysis} from subdivision meshes \cite{GasparGL09} to generic unstructured meshes. 
Once surface multigrid has become a reliable solver for linear systems
on manifold meshes, generalizing it to non-manifolds or point
clouds would be \update{another exciting} future direction. 

Another avenue for future work is to further optimize each component of the 
prolongation construction and multigrid solver routines.
\update{Although our method outputs a bijective map in most cases, bijectivity is not guaranteed. A more rigorous analysis is required to identify potential edge cases that may result in non-bijective maps.}
Currently, we use off-the-shelf simplification and distortion objectives
(as-rigid-as-possible \cite{LiuZXGG08} and conformal \cite{LevyPRM02} energies),
but these methods that are designed for other purposes may not be the optimal
ones for surface multigrid methods. 
For instance, we notice that the distortion in the self-parameterization is not
closely correlated to the convergence of our multigrid solver (see
\reffig{mg_decType}). We however use the off-the-shelf parameterization energy
designed to measure the distortion in our multigrid solver.
Developing simplification and parameterization methods tailored-made for
multigrid solver performance could further improve eventual solver speed.
%

\update{The relationship between multigrid convergence and bijectivity requires a deeper understanding. Although we empirically demonstrate the superior performance of our prolongation compared to other non-bijective prolongations, bijectivity is not required for a multigrid method to converge. In our construction, we even pay the price of high distortion to achieve bijectivity along the boundary (zoom in \reffig{SSP_visualization}). Thus, a deeper understanding of the connections between distortion, bijectivity, and multigrid convergence is important to reach optimal performance. }




\begin{acks}
  \update{Our research is funded in part by New Frontiers of Research Fund (NFRFE–201), German-Israeli Foundation for Scientific Research and Development (I-1339-407.6/2016), European Research Council (714776 OPREP), NSERC Discovery (RGPIN2017–05235, RGPAS–2017–507938), Israel Science Foundation (504/16), the Ontario Early Research Award program, the Canada Research Chairs Program, the Fields Centre for Quantitative Analysis and Modelling and gifts by Adobe Systems, Autodesk and MESH Inc. 
  We thank Uri Ascher, Tiantian Liu, and Irad Yavneh for sharing inspiring lessons on multigrid methods; Zhongshi Jiang for helps on experiements. We thank members of Dynamic Graphics Project at the University of Toronto; Michael Tao and Kazem Cheshmi for valuable discussions; Benjamin Chislett for proofreading; and John Hancock for the IT support. We thank all the artists for sharing 3D assets.}
\end{acks}

%% file: sections/appendix.tex
\appendix
%
%
\begin{figure}[h!]
  \centering
  \includegraphics[width=3.33in]{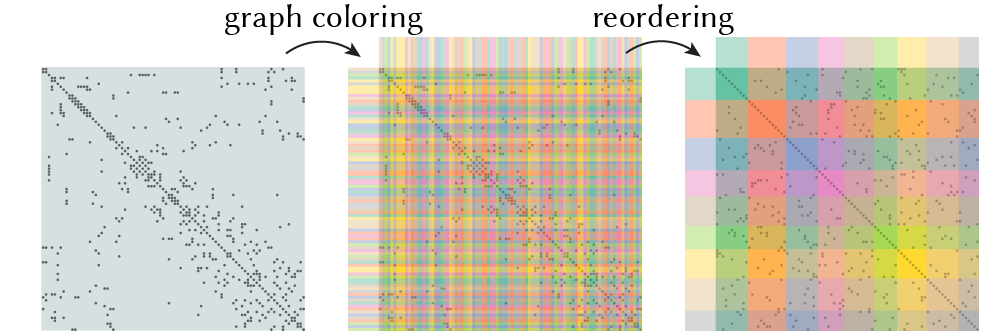}
  \caption{Given a system matrix with the sparsity pattern showing on the left,
  we \update{first} use a greedy graph coloring \update{approach} detailed in \refalg{graph_coloring} to
  ``paint'' the variables that are independent \update{of} each other with the same color
  (middle). Then we perform reordering to group the variables with the same
  color together (right) to parallelize our Gauss-Seidel relaxation.}
  \label{fig:graphColoring}
  \vspace{-5pt}
\end{figure}
\IncMargin{1em}
\begin{algorithm}[h!]
  \SetKwInOut{Input}{Input}
  \SetKwInOut{Output}{Output}
  \SetKwInOut{Parameter}{Param.}
    \caption{Graph Coloring}  
    \label{alg:graph_coloring}
    \Indentp{-1em}
    \Indentp{1em}
      \text{sort nodes $\{n_i\}$ by degree}\\
      $\textit{pallette} ← \{ \}$\\
      \For{each node $n_i$}{
        $N ← \text{gather colors from painted neighbors of $n_i$}$\\
        $c ← \text{find first entry in \textit{pallette} not occurring in $N$}$\\
        \If {$c$ is not found}
        {
          $c ← \text{new color}$\\
          $\text{append $c$ to \textit{pallette}}$\\
        }
        \Else{
          \text{move $c$ to back of \textit{pallette}}\\
        }
        \text{paint $n_i$ with color $c$}
      }
\end{algorithm}
\DecMargin{1em}
\section{Multi-Color Gauss Seidel}\label{app:graphColoring}
Our multigrid method spends a lot of the runtime on the Gauss-Seidel relaxation. We further accelerate the Gauss-Seidel relaxation by exploiting graph coloring (see \reffig{graphColoring}), a standard optimization.
Specifically, we treat the non-zero off-diagonal entries of a given sparse
matrix $\mA$ as a graph. We color this graph so that each node has a different
color from its neighbors using a simple modification of the method proposed by
\citet{WelshP67}, summarized in \refalg{graph_coloring} and repeated here for
completeness. We color each node in descending order of degree. When considering
node $i$, we try each color from a list of $k$ colors that have been previously
used for nodes $(1, \cdots, i-1)$. A color choice is valid if not matching any
of the previously colored neighbors of node $i$. If valid, node $i$ is colored
and that color is moved to the back of the list. If no valid color is found in
the list, a new color is used and added to the back of the list. This algorithm
has $O(|V| log |V| + |E| k )$ runtime and $O(|V| + |E|)$ memory complexity,
respectively, where $k$ is the number of output colors (for sparse matrices, $k
≪ |V|$). Although suboptimal (finding the optimal coloring is NP-complete), it
handily outperforms the method of \cite{FratarcangeliTP16} in runtime, memory
usage, and color parsimony. By moving selected colors dynamically to the back of
the list, we achieve better color balance (see, e.g., \reffig{graphColoring})
than considering the list in fixed order of insertion.

\section{Dirichlet Boundary Conditions}\label{app:Dirichlet}
Solving a linear system $\mA\vx = \vb$ is equivalent to minimizing a quadratic energy 
\begin{align}
  E(\vx) = \frac{1}{2} \vx^\top \mA \vx - \vx^\top \vb
\end{align}
where one can derive the same linear system by setting $\nicefrac{∂ E}{∂ \vx} = 0$. One way to handle Dirichlet boundary conditions $\vx(\textit{known}) = \vc$ is to reformulate the quadratic energy using only the unknown variables. Here we use \textit{known} and \textit{unknown} to represent indices of knowns and unknowns. We further use $\vx_k = \vx(\textit{known})$ to denote known variables and $\vx_u = \vx(\textit{unknown})$ for unknown variables in $\vx$. For matrices, we follow the same notation. For example, we use $\mA_{uk} = \mA(\textit{unknown}, \textit{known})$ to represent the corresponding sub-block in matrix $\mA$. We then rewrite the energy as (assuming $\mA$ is symmetric)
\begin{align} \label{equ:unknownOnly}
  E(\vx_u) = \frac{1}{2} \vx_u^\top \mA_{uu} \vx_u  + \vx_u^\top \mA_{uk} \vx_k - \vx_u^\top \vb_u + \text{constant}.
\end{align}
By setting the derivative to zero, we can derive a reduced linear system for only unknowns
\begin{align}
  \underbrace{\mA_{uu}}_\text{LHS} \vx_u  = \underbrace{- \mA_{uk} \vx_k + \vb_u}_\text{RHS}
\end{align}

We can leverage the same trick to incorporate Dirichlet constraints in the multigrid system. We use $\vx_c$ to denote the coarse variable such that $\vx = \mP \vx_c$ where the $\mP$ is the Galerkin prolongation operator. We can then write the unknowns as
\begin{align}
  \vx_u = \mP_{u:} \vx_c
\end{align}
where $\mP_{u:} = \mP(\textit{unknown}, :)$ (MATLAB notation) represents the rows of $\mP$ that correspond to the \textit{unknown} indices. Adding this to \refequ{unknownOnly} leads to
\begin{align} \label{equ:unknownOnly}
  E(\vx_c) &= \frac{1}{2} {(\mP_{u:} \vx_c)}^\top \mA_{uu} {(\mP_{u:} \vx_c)}  + {(\mP_{u:} \vx_c)}^\top \mA_{uk} \vx_k  \\ 
  &\qquad - {(\mP_{u:} \vx_c)}^\top\vb_u + \text{constant}. \\
  &=\frac{1}{2} \vx_c^\top \mP_{u:}^\top  \mA_{uu}\mP_{u:}\vx_c + \vx_c^\top \mP_{u:}^\top \mA_{uk} \vx_k \\ 
  &\qquad - \vx_c^\top \mP_{u:}^\top  \vb_u + \text{constant}. 
\end{align}
Similarly, setting the derivative with respect to $\vx_c$ results in 
\begin{align}
  \underbrace{\mP_{u:}^\top  \mA_{uu}\mP_{u:}}_\text{reduced LHS} \vx_c  = \underbrace{\mP_{u:}^\top (-\mA_{uk} \vx_k +\vb_u)}_\text{reduced RHS}
\end{align}
We can notice that, except at the second finest level where we need to extract the rows in prolongation that correspond to the \textit{unknowns} $\mP_{u:}$, we can solve the linear system at coarser levels without worrying about the constraints. 

Another special case may occur when there are too many \textit{known} indices. If too many variables in $\vx$ are given the reduced system $\mP_{u:}^\top  \mA_{uu}\mP_{u:}$ may have completely zero rows/columns. To handle this edge case, we further remove the columns on $\mP_{u:}$ where the maximum value is zero and the corresponding rows in the prolongation at the next level.

\section{Successive self-parameterization} \label{app:SSP}
Our pre-computation of multigrid hierarchy involves decimating the triangle mesh with successive self-parameterization and mapping vertices on the fine mesh to the coarse mesh to obtain their barycentric coordinates. We report the runtime of both pre-computation steps in \reffig{runtime}. 

Implementing successive self-parameterization only requires a small change to an existing edge collapse algorithm. Specifically, right after collapsing a single edge, the only modification is to use the method described in \refsec{jointFlattening} and \refsec{boundary} to formulate the joint variable and then flatten both patches to a consistent UV domain. To determine whether the collapse and the flattening is valid, we refer to the Appendix B in \cite{LiuKCAJ20} for more details.
During the querying stage, for a given query point represented as barycentric coordinates, we simply go through the list of local joint UV parameterization we stored from the decimation stage and update the barycentric coordinates successively using the method described in \reffig{baryQuery}. 
We pre-store the face indices involved in each edge collapse so that for each query point, we can easily check whether this point is involved in the collapse via checking the face indices.

\section{Ablation Study}\label{app:ablation}
\begin{figure}
  \centering
  \includegraphics[width=3.33in]{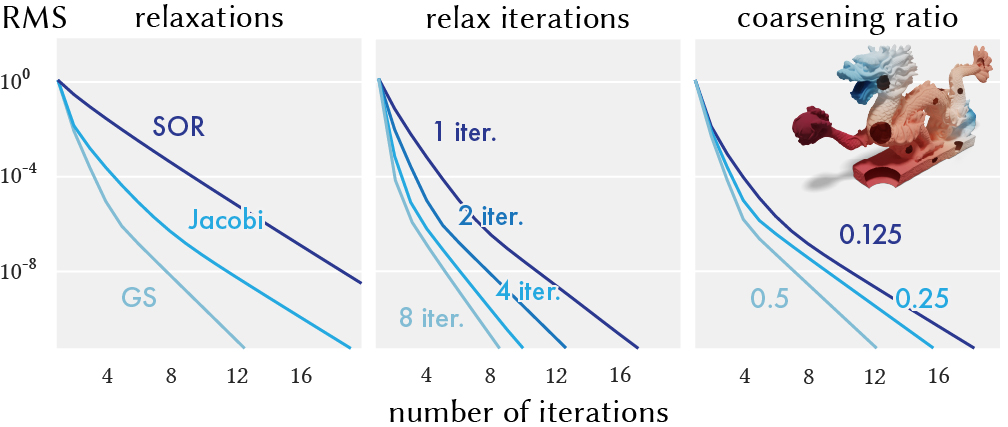}
  \caption{We conduct an ablation study on \update{the} multigrid hyperparameters, including the relaxation method (left), the number of relaxation iterations (middle), and the coarsening ratio (right). \textcopyright model by Oliver Laric under CC BY-NC-SA.}
  \label{fig:ablation}
  \vspace{-5pt}
\end{figure}

\update{In addition to} the prolongation operator, the hyperparameters of a multigrid method also play a role in the convergence behavior. 
In terms of stopping criteria, the accuracy $ε$ depends on the problem and the size of the mesh. We usually set $10^{-5} ≤ ε ≤ 10^{-}$ in order to get visually \update{indistinguishable} results compared to the ground truth. 
Using a reasonable initialization, such as the vertex positions in the previous iteration in mesh deformation, would further reduce the number of iterations to get the desired accuracy. 
For other hyperparameters, we conduct ablation studies on the choice of relaxation methods, pre-/post-relaxation iterations $\mu_\textit{pre},\ \mu_\textit{post}$, and the coarsening ratio between consecutive levels (see \reffig{ablation}).
In terms of the relaxation methods, Gauss-Seidel is usually the go-to choice due to its effectiveness in smoothing out the high-frequency error. \update{Practitioners} may also prefer the (damped) Jacobi because it is faster and easier to parallelize, even though each iteration is less effective. 
In terms of the number of relaxation iterations, usually a couple of iterations (2 or 3) are sufficient to handle the high-frequency error. While we also notice that some multigrid methods (e.g., \cite{xian2019scalable}) use lower-order prolongation with \update{many} more relaxation iterations to compensate for the inter-grid transfer error. 
In terms of coarsening ratio, using a less aggressive coarsening (e.g., 0.5) could reduce the error caused by the inter-grid transfer, but it often results in a bigger multigrid hierarchy and a longer runtime per cycle. On the other hand, using a more aggressive coarsening often leads to large inter-grid transfer error and slow convergence.
Our default setup coarsens the geometry down to 0.25 of its previous resolution until we reach the coarsest mesh with no lesser than 500 vertices. In each V-cycle, we use the Gauss-Seidel relaxation with 2 pre- and post-relaxation iterations. Our experiments suggest that the optimal set of parameters that minimizes the wall-clock runtime depends on the geometry and the PDE of interest. But we use our default parameters for all our experiments in \refsec{applications} for consistency. 

\section{Fast Data Smoothing}\label{app:fastDataSmoothing}
When we discretize the data smoothing energy \refequ{dataSmoothing}, we often arrive the following linear system
\begin{align}
  \underbrace{(α \mQ + (1-α) \mM)}_{\mA} \vx = (1-α) \mM \vf
\end{align}
where $\mQ$ is a matrix that depends on the choice of the smoothness energy, $\mM$ is the mass matrix, $\vf$ is the noisy function, and $\alpha$ is the smoothness parameter. In order to build the coarsened system matrix, a straightforward implementation would be doing $\mP^\top\mA\mP$ directly, but we can actually split the computation via
\begin{align}
  \mP^\top\mA\mP  =  α (\mP^\top\mQ\mP) + (1-α) (\mP^\top\mM\mP).
\end{align}
Then we can pre-compute $\mP^\top\mQ\mP$ and $\mP^\top\mM\mP$ even before knowing the parameter $\alpha$. As a results, during the online stage when a user \update{adjusts} $\alpha$, we only require an efficient matrix addition to \update{compute} the system matrices for all the multigrid levels.